\documentclass[apjl]{emulateapj}
\usepackage{epsfig}

\begin{document}
\title{On the Planet and the Disk of CoKuTau/4}
\author{Alice C. Quillen, Eric G. Blackman, Adam Frank, Peggy Varni\`ere}
\affil{Department of Physics and Astronomy, University of Rochester, Rochester, NY 14627}
\email{aquillen, afrank, blackman, pvarni @pas.rochester.edu}
\begin{abstract}
Spitzer observations of the young star CoKuTau/4 reveal a disk
with a 10 AU hole that is most likely caused by a newly formed planet.
Assuming that the planet opened a gap in the viscous disk, 
we estimate that the planet mass is greater than 0.1 Jupiter masses. 
This estimate depends on a lower limit to the
disk viscosity derived from the time scale needed to accrete the
inner disk, creating the now detectable hole. 
The planet migration time scale must at least modestly exceed
the time for  the spectrally inferred hole to clear.
The proximity of the
planet to the disk edge implied by our limits
suggests that the latter is perturbed
by the nearby planet and may exhibit a spiral pattern rotating with
the planet. This pattern might be resolved 
with current ground based mid-infrared cameras and optical cameras
on the Hubble Space Telescope. 
The required sub-Myr planet formation may challenge 
core accretion formation models. However, we find that only if 
the planet mass is larger than about $10$ Jupiter masses, 
allowing for a high enough surface density without inducing migration,
would formation by direct gravitational instability be possible.

\end{abstract}

\keywords{
stars: individual (CoKuTau/4) --- 
stars: planetary systems --- 
planetary systems: protoplanetary disks --- planetary systems: formation
} 

\section{Introduction}
There remains considerable debate about the nature and time scales
associated with planet formation in the disks surrounding young
stars \citep{boss02,mayer,pollack}.  The recent discovery of a
young star CoKuTau/4  (age $\tau_{age} \approx 10^6$ years)  
with a disk containing a 10 AU hole \citep{forrest,dalessio} suggests that
planet formation can take place quite early in the evolution of
protostellar systems. The critical theoretical link between the
inner hole in the disk and the presence of a planet is the
planet's ability to open a gap which impedes mass flow to smaller
radii.  Disks with edges and corresponding
inner cleared regions have been seen in older objects on large
scales such as HR 4796A, with a disk edge at $\sim 65$ AU
\citet{schneider} and GM Aur with a disk edge at $\sim 300$ AU,
\citep{rice}. However the recent Spitzer observations for smaller
scale hole in the younger CoKuTau/4 systems provide new insights
in the process which form planets and the evolution of planet-disk
systems.  
CoKuTau/4's age and the presence of the inner disk hole place
limits on time scales for the disk-planet
interaction, and constrain the properties of the
disk and planet. As we will show, the resulting estimates support the plausibility of the planet-disk scenario for CoKuTau/4.

Because the outer disk is still present and has a spectral energy
distribution similar to other T-Tauri stars lacking inner holes,
we suspect that the outer disk is still accreting. Massive planet
formation is unlikely to occur at extremely small radii from the
star, so following the formation of a planet, accretion of the
inner disk would leave a hole.   Without the planet
holding the accretion disk back, the disk edge would not be as
sharp as implied by the spectral energy distribution
\citep{dalessio}. For there to be an observable hole, 
the viscous time scale, $\tau_{\nu}$, at the outer edge of the hole 
must be less than the age of the system.  
That the planet has not yet migrated to smaller radii 
and the presence of an inner hole
allow us to constrain the properties of the planet and disk and 
estimate the probability of finding such systems.

\section{Clearing the Inner Disk and Constraints on Disk Viscosity}

We parametrize the disk viscosity, $\nu$, with a Reynolds number,
${\cal R} \equiv r^2 \Omega/\nu$ where $\Omega$ is the angular
rotation rate of a particle in a circular orbit around the star
and $r$ is the radius. Using a viscosity $\nu=\alpha c_s h$, where
$\alpha$ is the viscosity parameter, $c_s$ is the sound speed and
$h$ is the vertical disk scale height, we have
\begin{equation}
{\cal R } = \alpha^{-1} \left( { v_c \over c_s } \right)^2 =
\alpha^{-1} \left( { r \over h }\right)^2
\end{equation}
where $v_c = r \Omega$ is the velocity of a particle in a circular
orbit.  The time scale for the disk to accrete inward is $\tau_\nu
\sim {\cal R} \tau_{orb}/2\pi$ where $\tau_{orb}$ is the orbital
period.
To estimate $\tau_\nu$ we require the
disk aspect ratio $h/r$. From hydrostatic equilibrium, we have
$h/r \sim c_s/v_c$.
The sound speed is given by $c_s^2\sim {k_B T
/ m_p}$ where $k_B$ is Boltzmann's constant, and $m_p$ is the
proton mass. The temperature $T$, if set by radiative balance
with the star, is $T \sim \left({ L_\odot  \over 16 \pi r^2
\sigma}\right)^{1/4}$, where $\sigma$ is the Stefan-Boltzmann
constant. Heat released in the disk by accretion 
could lead to a higher temperature than that estimated above. Alternatively,
the disk could be self-shielding, cooling the mid-plane 
below the temperature estimated above (e.g.,
\citealt{chiang}). Keeping these complications in mind we use this
temperature as a starting point.

CoKuTau/4 is an M1.5 star at a distance of 140pc with an estimated
mass of $0.5 M_\odot$, $\tau_{age}=10^6$  years,  
and luminosity of $0.6 L_\odot$ (\citep{dalessio,KH95}).
Assuming an inner edge of 10 AU and the luminosity and
mass given above, at the disk edge we find $T \sim 80K$, $c_s \sim
0.8$km/s and $v_c \sim 7$km/s. The orbital period at 10 AU is $\sim
40$ years. Using the this sound speed and 
hydrostatic equilibrium, $h/r \sim 0.1$.  However, if the disk
is optically thick, $h/r$ could be lower.
We use $h/r = 0.05$ below, though this particular
choice is not significant.

If a planet
exists in CoKuTau/4's disk at $r = $10 AU, then the hole formed
because disk material is prevented from accreting across the
planet's orbital radius by the transfer of orbital angular
momentum from the planet to the disk. The presence of the hole
therefore implies that the material in the inner disk 
($r < $10 AU) has had time to accrete onto the star. This requires
$\tau_\nu < \tau_{age}\simeq 1$Myr, which in turn implies
\begin{equation}
\tau_\nu = {\cal R} \tau_{orb}/2\pi = \alpha^{-1} \left({r \over
h} \right)^2 \tau_{orb}/2\pi < 1 {\rm Myr}.
\end{equation}

Using $\tau_{orb} =$40 years, the above inequality implies that
${\cal R} \lesssim 1.6\times 10^5$. Using  $h/r = 0.1$, we also
require that $\alpha \gtrsim 6\times 10^{-4}$. These values
are within theoretical expectations.  A viscosity
parameter $\alpha \sim 0.01$ typically emerges from protostellar disk model
fitting \citep{Hartmann}.  Note that if $\alpha$ or $h$ were
orders of magnitude lower, the inner disk would not have had time
to accrete onto the star. In principle, the viscosity of the inner
disk could differ from that of the outer disk, or either could
have dropped in the past million years, but here we are just
assuming the simplest steady case. The result highlights that the
hypothesis that the inner disk has accreted onto the central star
is consistent with age constraints and reasonable disk properties.
If instead we use $\alpha \approx 0.01$ and $h/r = 0.05$ we find
${\cal R} \sim 4 \times 10^4$ and 
$\tau_{\nu} \approx 2\times10^5$yr, also less than $\tau_{age}$
and thus in principle consistent with the observations.
However, the planet formation time  must
be less than the viscous time at the radius where the
planet forms. A viscous time significantly less
than 1 Myr challenges core accretion models if the planet
formed at the corresponding radius.

Instead of accreting onto the central star, the inner
disk could have been depleted by multiple planet formation,
or agglomeration into large dust grains and planetesimals 
within 10 AU. Here however we focus on  the simplest paradigm of a single 
planet and viscous accretion.

\section{Gap Opening and Constraints on Planet Mass}
The inner disk will begin accreting onto the star once a newly
formed planet opens up a gap.  The condition for opening a gap
provides a limit on the planet's mass. Without a gap, 
the disk would  accrete unimpeded through  the orbital radius
of the planet. A gap decouples the inner
and outer disk, except through their
interaction with the planet. Because the planet acts as a time
dependent gravitational potential perturbation, 
it can resonantly drive waves
at Lindblad resonances into a the gaseous outer  planetesimal disk
\citep{GT78,lin79,ward}. These waves carry angular momentum
and therefore govern both how a planet opens gaps  (e.g.,
\citealt{bryden,artymowicz94}) as well as the the radial migration
(e.g., \citealt{nelson,ward}).

To open a gap,  a planet must be
sufficiently massive that spiral density waves dissipated in the
disk overcome the inward flow due to viscosity (\citep{lin79,bryden,ward}).  
This leads to the
condition \citep{nelson}
\begin{equation} q \gtrsim 40 {\cal R}^{-1}, 
\end{equation} 
where $q \equiv M_p/M_*$, the mass of the
planet divided by that of the star.  For ${\cal R} \lesssim 1.6
\times 10^5$ estimated above for CoKuTau/4, the gap opening
condition implies that $q \gtrsim 2.5\times 10^{-4}$, or $M_p \gtrsim
0.1  M_J$ where $M_J$ is the mass of Jupiter.

\section{Migration, Mass Loss and Surface Density}
Once formed, a planet's interaction with a surrounding disk may
lead to the transfer of angular momentum and the migration of the
planet closer to the star \citep{ward}. 
For more massive planets,
migration can occur after a gap is opened (denoted Type II migration). 
The condition for opening a gap and the nature of Type II
migration are linked.  A gap is maintained when the torque density
from spiral waves driven at different resonances balances the
inward torque due to viscous accretion. If the planet mass is less
than or comparable to the disk mass with which it interacts then
the planet will migrate on viscous time scale; it behaves as
``just another particle'' in disk 
and $\tau_{mig} \sim \tau_\nu$ (e.g., \citealt{nelson}).
However this result is subject to the surface density
profile; outward migration rather than inward migration may occur 
\citep{masset03}.

If the planet mass is large compared with the disk mass with which
it interacts then the inertia of the planet slows migration and
$\tau_{mig} > \tau_{\nu}$.  That CoKuTau/4's planet still
resides a large distance from the star implies that significant
migration has yet to occur. The lack of significant migration
along with the presence of a hole implies 
$\tau_{clear}< \tau_{mig}$, where $\tau_{clear}$ is
the timescale for the gas within the planet's semi-major
axis to accrete onto the star.  If we use $\tau_{clear}
\sim \tau_{\nu}$ then we require $\tau_{mig} \ge \tau_{\nu}$ for type II migration.

After a gap is opened, the inner disk accretes onto the star while
outer disk experiences a pile up of material at the edge of the
disk exterior  to the planet.  Once $M_{edge} \sim M_p$, migration
would take place on a viscous time scale. Here $M_{edge} = \pi r^2
\Sigma_e$ where $\Sigma_e$ is the 
surface density just outside the
disk edge. The accretion rate of the outer disk  may be
crudely estimated to be
\begin{equation}
\dot{M}_a \sim {M_{edge}\over \tau_\nu} \sim  { M_{p}\over
\tau_{\nu}} \sim 
 10^{-9} M_\odot{\rm yr}^{-1}
\left({M_p \over 1 M_J} \right) 
\left({10^6 {\rm yr} \over \tau_\nu } \right).
\end{equation}
Significantly larger accretion rates  would have led to larger
accumulations of mass in the disk edge and an earlier onset of
migration.  We note that accretion rates in the range of value of
$\dot{M}_a = 10^{-7} - 10^{-10} ~M_\odot/$yr are consistent with
observations of accretion rates in million year old evolved
T-Tauri systems \citep{calvetpp4}. Larger  values of $\dot{M}_a$
could be accommodated in our calculations by taking a
larger planet mass planet and/or a larger required mass in the
disk edge to initiate migration. 
If we assume that the build up of $M_{edge}> M_p$ would have led to inward
migration, its absence constrains the disk surface
density through $M_p \gtrsim \pi r_p^2 \Sigma_e$,
where $r_p$ is the semi-major axis of the planet. Using 
our limit $M_p > 0.1 M_J$ estimated above, we find
$\Sigma_e \le 4$ gm~cm$^{-2}$ in the disk edge,
a plausible value for disks around young stars.

The discovery of a planet at $\sim 10AU$ from
CoKuTau/4 is therefore consistent with the hypothesis 
that inward disk migration has not proceeded to completion
because sufficient mass has not
accumulated in the disk edge.

\section{Predicted disk morphology}
Having established that a young 
sub-Jovian mass planet orbiting at 10 AU in  the CoKuTau/4 disk is plausible, 
we now consider the
gravitational and hydrodynamic interaction between the planet and
disk. Using the hydrodynamics code developed by
\citet{masset00,masset02} we have performed a 2D hydro
simulation using the parameters estimated in the previous sections.
Figure 1 shows the morphology from a simulation with 
planet mass ratio
$q = 3 \times 10^{-4}$, Reynolds number ${\cal R} = 10^5$,
and disk aspect ratio $h/r = 0.05$. In the simulation,  the planet
was initially set into a circular orbit, with the disk edge
located at 1.1 times the semi-major axis of the planet's orbit.
The initial surface density was taken to be $\Sigma =\Sigma_0
(1.1r_{p}/r)^{-1}$ where $\Sigma_0= 10^{-4}M_*/r_p^2$ is the
surface density at the disk edge and $r_p$ is the planet's
semi-major axis. For $r<1.1r_p$ we set the disk density to be 100
times lower than that at the disk edge to approximate an initial
inner hole. The planet is free to migrate via gravitationally
induced angular momentum exchange with the disk, and can accrete
gas within its Roche Lobe. Note that the simulated disk is not
massive enough for self-gravity to play a role. Figure 1 shows the
gas density at time $t=100\tau_{orb}$ after the beginning of the
simulation.

The principle conclusion from our simulations relates to
spiral density waves driven into the disk from interactions with
the planet.  Because of the proximity of the planet to the disk
edge, the disk contains more than one Lindblad resonance. Multiple
spiral density waves can be driven in the disk edge by the planet
at these resonances. The range in radius where the waves are
launched is of order the scale height, $h$ \citep{artymowicz}.
Consequently we expect the winding of the spiral pattern to depend
on the scale height and hence on the disk temperature, i.e. the
spiral wave would be more tightly wound if the disk were cooler
and the scale height smaller. The two-armed pattern rotates with
the planet and is probably caused by the combined effect 
of more than one density wave, as
explained by \citep{ogilvie}.  
The spiral pattern could be observable by high
angular resolution imaging.  At a distance of 140pc, $0.1'' \sim
14$ AU.  The features exhibited in the simulation are then close to the
resolution limit of ground based 10-meter class telescope at $10\mu$m
or by the Hubble Space Telescope (HST) at optical wavelengths.
Thus it is possible that the spiral pattern might also be detected
in optical scattered light images.  In addition, the 
asymmetry of the disk hole caused by the arm which extends toward the planet
may be detectable through imaging. 

When the Reynolds number or the planet mass is higher, the disk edge
would be further away from the planet. Then the lower $m$ Lindblad
Resonances are the dominant sites of density wave driving.
Consequently, 2 or 3 density peaks, corresponding to the
constructive interference of 2 and 3 armed waves might be detected
in images of the disk edge.   If the disk aspect ratio is higher
than considered here, a one-armed pattern rotating
with the planet dominates over the two-armed one seen in Figure 1.

\vskip 0.05truein 
{\centering
\includegraphics[angle=0,width=3.3in]{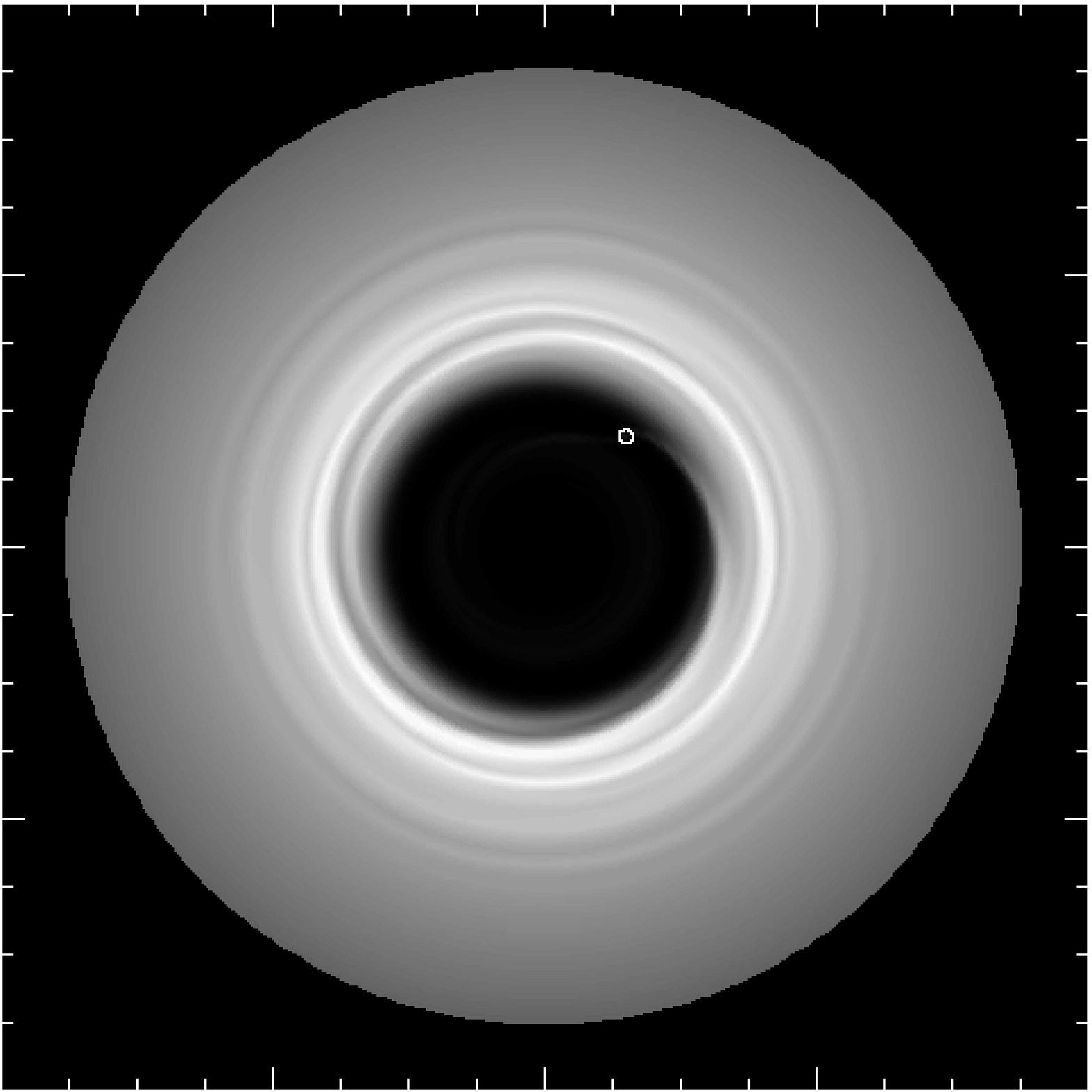}
}
\begin{quote}
\baselineskip3pt {\footnotesize Fig.~1-- 
Morphology of a planet
disk system with  planet mass 
of 0.1 Jupiter mass for the CoKuTau/4 system. The disk aspect
ratio is $h/r = 0.05$ and Reynolds number ${\cal R}= 10^5$. Because
of the proximity of the planet and disk, high order spiral density
waves are driven by planet planet at Lindblad resonances. If these
waves constructively interfere, the result is a spiral
pattern that rotates with the planet.
The planet in this simulation is accreting gas.}
\end{quote}

Hydrodynamic simulations have shown differences between 2D and 3D
models \citep{makita}, particularly in the opening angle of the
spiral density waves. In a 3D
disk, the waves may not constructively interfere as they do in our
2D simulation, because of differences in the dispersion relation
\citep{ogilvie}.  Further 3-D simulations are required to better determine
structures in the disk formed via planet-disk interactions.

In the simulation shown, we allow the planet to accrete 80\% of
the material found within its Roche Lobe.   The remaining 20\% can
flow past the planet into the inner disk (as seen
previously by \citealt{lubow99}), accounting for the
non-zero gas density within the planet's semi-major axis. 
If the disk has the high viscosity considered here, then it is likely
that the planet is still accreting significantly. 
A planet that is still accreting may be
surrounded by a hot observable circumplanetary disk (e.g., \citealt{lubow99}).  

The Spitzer observations imply that there 
is very little dust within the disk edge \citep{dalessio},
though the presence of larger bodies is less constrained. 
Better hydrodynamic modeling will help to understand  the flow 
past the planet into the inner hole and consequently on
the disk, planet mass and planetary accretion.

\section{Discussion and Conclusions}

The recent discovery by \citet{forrest} of a young system with 
a 10 AU hole \citep{dalessio} allows us to explore new constraints on
the evolution of young planets and circumstellar disks. Based on
the assumption that the inner disk has accreted onto the star
CoKuTau/4 within a time equivalent to the age of the star we
estimated that  ${\cal R}\lesssim 1.5 \times 10^5$.
Using this, we evaluated the condition for a planet
to open a gap in the disk and  found that the planet within the disk
edge of CoKuTau/4 is greater than $0.1M_J$.
The apparent lack of inward
migration of the planet leads to estimates of the disk accretion
rate and surface density which are consistent with observations of
evolved T-Tauri systems.  Given the inferred planet mass, we
expect the disk edge to be very near
the planet. This implies that the planet could be accreting
material and would interact strongly with the disk by driving waves into
the disk from resonances. Idealized simulations suggest that this
may produce a two-armed spiral pattern, rotating with the planet
and extending a few disk scale heights away from the radius of the planet.
Structure in the disk could be detectable with
$0.1"$ high resolution imaging in scattered optical light with HST
or by ground based 10m class telescopes in the mid-infrared.

The contemporaneous presence of both a sharp edge and an inner
hole (suggestive of viscous inner disk clearing) 
implies that the hole clearing timescale is
less than the planet migration timescale, $\tau_{clear}<\tau_{mig}$, 
otherwise, the planet would have disappeared along with the hole, and the
hole re-filled.  
The probability to see both an edge from the
planet induced gap AND an disk hole for $r<(10AU-\delta r_{gap}/2)$ is then 
\begin{equation}
P=[\tau_{mig}(10 AU) - \tau_{\nu}(10 AU- \delta r_{gap}/2)]/\tau_{age},
\label{5}
\end{equation}
where $\delta r_{gap}$ is the gap width opened initially by the 
planet and the time scales are meant to
be taken at the radius in parentheses. 
For $\delta r_{gap}<< 10AU$,  
$P\sim (\tau_{mig}(10AU)-\tau_{vis}(10AU))/\tau_{age}$.
Once the planet appears at 10AU, it takes at least one viscous timescale for
material to pile at the disk edge, and presumably 1 more for
migration to take place, so then $P\sim \tau_\nu/\tau_{age}$.
For $\alpha=0.01$ the estimate for $\tau_\nu$ of section 2 would then give 
$P\sim 10$\%. This argument presumes that  planet appearance at $10AU$
occurs  before a viscous time at $10AU$.
(This is more restrictive than the minimal condition
that a planet must form faster than a viscous time scale at its
formation radius because the planet could have formed at $r > 10AU$.)

More robust processes to explain why the inward 
migration time  scale $\tau_{mig}<\tau_{clear}$
are possible and may be necessary.  \citet{masset03} find that
surface density profiles shallower than $r^{-1/2}$
can induce an outward planet migration rather than inward
which could certainly explain the presence of wall and disk.
 Alternatively, if the inner accretion disk had a
radially dependent viscosity coefficient, or incurred a change of
accretion mode (e.g. non-viscous transport) the surface density
profile might evolve so as to produce a hole within the planet
radius even if the planet migrated on a local viscous time.  More
work on these possibilities for CoKuTau/4 will be needed.

Finally, we note that the planet-disk scenario for CoKuTau/4
implies that planets of greater than two Neptune masses can form in $\le 10^6$
years. This may push the envelope of core-accretion models
\citep{pollack} and this may appear to
favor planet formation by direct gravitational instability
\citep{boss02,boss03,mayer}. Our results  constrain
the properties of the disk from which the planet formed:
The Toomre instability parameter for a thermally supported disk at
the planet orbit radius $r_p$ is $Q = 200 \left({M_*\over
0.5M_\odot}\right) \left({\Sigma\over 4 {\rm g/cm^2}}\right)^{-1}
\left({r_p\over 10 AU}\right)^{-2} \left({h/r_p\over 0.05}\right)$,
where we have scaled to to the values based on our lower
mass limit for the planet in CoKuTau/4. Planet formation by direct
gravitational instability is only possible for $Q\lesssim 1.5$.
For this condition to be satisfied, the 
disk would have had be remarkably thin and/or have a very high $\Sigma$
in the not-to distant past.
Our upper limit on $\Sigma$ is proportional to $M_p$,
so that only if $M_p=10M_J$, would the upper limit of $\Sigma\le 400$g/cm$^2$,
allow $Q$ to be low enough for planet formation by
direct gravitational instability. Observational constraints on the surface
density are needed.


\acknowledgments
We thank Paula D'Alessio, Lee Hartmann, Nuria Calvet,
and the other members of the IRS-disks team for
sharing their work on the modeling of CoKuTau/4's spectral
energy distribution in advance of publication,
and Lee Hartmann for additional comments.
We thank Joel Green, Dan Watson, Judy Pipher, Mike Jura, Bill
Forrest and the IRS-disk team for helpful discussions.
Support for this work was
provided by NSF grants AST-9702484, AST-0098442, NASA grant
NAG5-8428, DOE grant DE-FG02-00ER54600,
and the Laboratory for Laser Energetics. 
This research was supported in part by
the National Science Foundation to the KITP under Grant No.~PHY99-07949.

\end{document}